\documentclass[acmsmall,natbib=false]{acmart}

\AtBeginDocument{%
  }

\setcopyright{acmcopyright} 
\copyrightyear{2024}
\acmYear{2024}
\acmDOI{XXXXXXX.XXXXXXX}

\RequirePackage{biblatex}
\begin{document}

\title{Scalability limitations of Kademlia DHTs when enabling Data Availability Sampling in Ethereum}

\author{Mikel Cortes-Goicoechea}
\email{mikel.cortes@bsc.es}
\orcid{0000-0003-3167-6014}
\authornotemark[1]
\affiliation{%
  \institution{Barcelona Supercomputing Center}
  \city{Barcelona}
  \country{Spain}
}

\author{Csaba Kiraly}
\affiliation{
  \institution{Status.im}
  \country{Italy}}
\email{csaba@status.im}

\author{Dmitriy Ryajov}
\affiliation{%
  \institution{Status.im}
  \country{Costa Rica}}
\email{dryajov@status.im}

\author{Jose Luis Muñoz-Tapia}
\affiliation{%
  \institution{U. Politécnica de Catalunya}
  \country{Spain}}
\email{jose.luis.munoz@upc.edu}

\author{Leonardo Bautista-Gomez}
\affiliation{%
 \institution{Status.im}
 \country{Spain}}
\email{leo@status.im}

\renewcommand{\shortauthors}{Cortes et al.}

\begin{abstract}
  Scalability in blockchain remains a significant challenge, especially when prioritizing decentralization and security. The Ethereum community has proposed comprehensive data-sharding techniques to overcome storage, computational, and network processing limitations. In this context, the propagation and availability of large blocks become the subject of research to achieve scalable data-sharding. This paper provides insights after exploring the usage of a Kademlia-based DHT to enable Data Availability Sampling (DAS) in Ethereum. It presents a DAS-DHT simulator to study this problem and validates the results of the simulator with experiments in a real DHT network, IPFS. Our results help us understand what parts of DAS can be achieved based on existing Kademlia DHT solutions and which ones cannot. We discuss the limitations of DHT solutions and discuss other alternatives.
\end{abstract}



\received{10 February 2024}

\maketitle

\section{Introduction}
\label{sec:introduction}
Scalability has proven to be one of the major challenges in blockchain technology, at least when trying to keep it decentralized and secure \cite{zhou2020solutions}, and Ethereum \cite{eth-whitepaper} is not different from the rest in that respect \cite{chauhan2018blockchain}. To tackle those scalability limitations, Ethereum plans to rely on the so-called Layer 2 (L2) solutions~\cite{layer2s}
~\cite{scaling-eth-layer2}, where we can find solutions such as private payment channels~\cite{papadis2020blockchain} or the recently popular roll-ups~\cite{thibault2022blockchain}. 
These latest protocols, which include mature projects such as Polygon\footnote{https://polygon.technology/papers/pol-whitepaper}, Arbitrum\cite{kalodner2018arbitrum}, Optimism\footnote{https://www.optimism.io/}, or zkSync\footnote{https://zksync.io/}, benefit from the speed of processing transactions off-chain. 
L2 solutions optimize their operation costs by aggregating multiple transactions into smaller ones, providing, in some cases, a set of proofs or commitments that can be used to verify the correctness of the operations. However, all of them still rely on the consensus and security of layer 1s to ensure that the interactions are traceable and immutable. 

Among the available solutions, roll-ups have been popularised over the last years for their similar yet cheaper EVM-compatible solutions. However, these protocols generally need L1's chain space to store the commitments of the processed data. Ethereum researchers have proposed data-sharding, colloquially known as ``DankSharding"~\cite{danksharding}, to facilitate that needed space for roll-ups. The idea is to offer larger, cheaper, and off-chain block space (blobs) to fill up with verifiable commitments data\footnote{https://domothy.com/blobspace/}. The nature of roll-ups generally relies on fault or verifiable proofs as a mechanism to validate transactions.
This means that every time roll-up submits a bunch of transactions to the main chain, also sharing the respective verification data as a blob, the rest of the network will have $X$ number of weeks to fetch and verify if something went wrong. Under this premise, transactions could be verified as one honest verifier exists on the network during those weeks. 
But those proofs are not needed in the long term~\cite{eip4844}. Therefore, blob data can be considered ephemeral and discarded after a certain time~\cite{protodanksharding}.
This protocol proposal will scale the execution payload size, as after the execution payload of a block, the related ``blobs" will also be shared on separate GossipSub~\cite{vyzovitis2020gossipsub} message broadcasting topics. This makes the total $block + blob space$ increase from the current average of $150KB$\footnote{https://etherscan.io/chart/blocksize, https://ethresear.ch/t/big-block-diffusion-and-organic-big-blocks-on-ethereum/17346} to something in the order of $32MB$.

The proposal relies on Erasure Coding \cite{wicker1999reed} and Data Availability Sampling (DAS) \cite{hall2023foundations} techniques to split the blob block into segments/chunks/samples to reduce the bandwidth, storage and computational overhead that process all blobs imply for nodes and validators.
Splitting and sharing the blob block in smaller pieces makes the propagation of blobs easier through the network, as the validators will only have to subscribe, receive, and process a smaller portion of the larger coming blob block. Ordinary (non-validator) nodes instead rely entirely on probabilistic sampling, receiving only a few segments of the block to verify that the block is reconstructable using the erasure code with very high probability.
The challenge of increasing the block size from an average of $150KB$ to $32MB$ is not simple. The upgrade to allocate more blobs in a cheaper ephemeral block space has many configurations and parameters that must be adjusted to make it ``feasible". Figure \ref{fig:das-block-split} represents the proposed Execution Layer block's division for DAS, where the block will be first split into $256$ rows and columns and then equally extended in both directions (rows and columns) applying Reed-Solomon Encoding \cite{wicker1999reed} to ensure each row and column (and therefore the entire block) can be reconstructed as long as half of extended row or column samples are retrievable. 

Due to the synergies between the existing blob samples' addressing needs from Ethereum's DAS and the content addressing capabilities of DHTs, this paper presents the results of simulating a Kademlia-based DHT for Ethereum’s data-sharding case. The paper explores the time restrictions of common DHT operations under the expected DankSharding workload. We present a configurable DHT simulator that can reproduce the DHT operations under connection errors and network latencies, essential to recreate not that foreseeable peer-to-peer networks accurately. Furthermore, We raise concerns about the scalability limitations of DHTs like IPFS's when they suffer a high throughput of store and retrieval operations. Especially when the protocol wants to keep the hardware resources close to commodity home hardware as Ethereum does \cite{cortes2021resource} to promote decentralisation by allowing home node operators. Considering the outcome of this research, we propose a set of alternatives that could potentially overcome the listed limitations, ensuring low overhead for the nodes while still providing the expected requirements for blockchain scalability. 

The remainder of this paper is organized as follows. Section~\ref{sec:related-work} introduces in further detail what makes Kademlia DHT a suitable candidate for Ethereum's DAS.
Section~\ref{sec:related-work} discusses related work done on high throughput DHTs and previous DAS attempts in the ecosystem.
Section~\ref{sec:methodology} introduces the methodology followed to simulate the DHT behaviour under Ethreum’s DAS proposal.
In Section~\ref{sec:evaluation}, we introduce and analyze the results obtained by our study, comparing our simulated results with live DHT networks and showcasing the existing limitations of DHTs under heavy usage.
Finally, Section~\ref{sec:conclusion} concludes the findings of this work and presents some future directions.

\section{How does a DHT fit in Ethereum's DAS approach?} 
\label{sec:why-dht?}

The challenge of increasing the block size from an average of $150KB$ to $32MB$ is not simple. The upgrade to allocate more blobs in a cheaper ephemeral block space has many configurations and parameters that must be adjusted to make it ``feasible". Figure \ref{fig:das-block-split} represents the proposed Execution Layer block's division for DAS, where the block will be first split into $256$ rows and columns and then equally extended in both directions (rows and columns) applying Reed-Solomon Encoding \cite{wicker1999reed} to ensure each row and column (and therefore the entire block) can be reconstructed as long as half of extended row or column samples are retrievable. 

With the current count of over $880.000$ active validators\footnote{https://ethseer.io/?network=mainnet} distributed around $12.000$ active Beacon Nodes\footnote{https://monitoreth.io/nodes\#act-nodes} (at the moment of writing this paper), DAS aims to disseminate the block by entire rows and columns through the usage of distinct GossipSub topics\cite{vyzovitis2020gossipsub}. It has been determined that each active validator will have to attest to a set of two random rows and columns in a single slot in an epoch \cite{cortes2023autopsy}. This will ensure optimal bandwidth-efficient block propagation. However, since part of the network's privacy relies on not knowing which node in the network hosts which validator (preserving the anonymity of the validators), it complicates checking which sample of the blob block is available on which node without compromising the privacy of the validators, as the topic subscription of each node could be correlated with the identity of validators.

In this privacy-preserving case, the proposed DAS schema still needs a resilient and privacy-preserving method able to route any blob verifier to the node that keeps it in the network, and this is where a popular DHT such as Kademlia enter into place.
The Kademlia DHT has been widely used for content addressing by many networks such as IPFS~\cite{benet2014ipfs}, where it has proven to ensure resilience against the constantly changing network while still providing fast content retrieval times and some privacy-enhancing solutions like ``double hashing"~\cite{ipfs-double-hashing} of the stored records. 
With all these key features, a Kademlia DHT is a suitable candidate to provide available samples to any interested node, as the seeding of the DHT reduces considerably the possibility of unveiling the location of validators. In this case, instead of seeding the DHT with the ``Provider Records" (pointers to the content provider in IPFS), because the size of each sample (as introduced in Figure \ref{fig:das-block-split}) is relatively small ($512$ bytes of data plus $48$ bytes of KZG proofs \cite{zkg-commitments}), we propose using a Kademlia-based DHT to evenly distribute the samples based on the Beacon Node's ID and the segment's distance. This would reduce the number of operations to retrieve each sample from two: i) ask the network who has the content, and ii) contact and download the sample from the node, directly to the second step. 
However, due to the highly demanding DAS approach of DankSharding, the limits of the Kademlia DHT still remain unknown.

\section{Related work}
\label{sec:related-work}

The overhead of scaling up the processing capabilities of modern blockchain can only be handled by splitting the workload across the participating actors. In that context of task division, preventing data withholding attacks and providing means to verify that the data had been released probabilistically is what Data Availability Sampling aims to solve. 
However, DAS is not a new concept in the blockchain ecosystem. Some prior works have attempted to bring DAS \cite{sheng2021aced} proofs on-chain through the usage of Oracles \cite{al2020trustworthy}. However, the idea relies on a third trusted blockchain to submit the data availability proofs. Thus, the idea cannot be directly applied to a network such as Ethereum, as it is one of the networks that generates trust for others.
Other projects explored different solutions for DAS. Celestia \cite{celestia} proposed a network with the single purpose of ordering and guaranteeing data availability \cite{al2019lazyledger}, splitting the network into three main actors: consensus nodes, storage nodes and light clients. Despite achieving certain scalability results, the proposal only ensures that light clients don't need all the storage to validate chain transactions. This heavily relies on a low and limited number of ``super nodes" that propose and download large blocks while serving all the segments to light clients. Due to their heavy hardware requirements, the approach exposes the network to a critical failure if storage nodes are overwhelmed by the workload. 

DHTs have been part of the peer-to-peer and web3 ecosystem for a while due to their censorship resistance capabilities, workload balancing properties, and logarithmically scaling performance. Many projects like IPFS \cite{benet2014ipfs} \cite{trautwein2022design}, Libp2p \cite{libp2p} or even Ethereum \cite{discv5} have relied on the Kademlia DHT implementation \cite{maymounkov2002kademlia} for content routing or peer discovery. Some works like \cite{dimakis2010network} have shown that despite needing some replication factor to overcome node churn in the network, the combination of erasure coding techniques and DHTs can reduce the bandwidth for the retrievability of the items. Since both techniques are robust and well-tested in the literature, one might expect that DHTs are a suitable solution to enable DAS on Ethereum. However, whether DHTs can guarantee the high throughput that DAS requires is uncertain.
Previous attempts have been to improve the latency and increase the throughput of DHTs. Previous work like \cite{dabek2004designing} presented viable solutions that help increase the DHTs' throughput while providing congestion control and minimizing the latency between nodes. However, the authors do not contemplate the extra overhead that connecting and publishing into a larger network like the Ethereum one could have, extracting the available throughput from simulations with under $500$ nodes.
This paper explores whether a modern Kademlia-based DHT can enable high-throughput native DAS techniques in a blockchain network. It focuses on a DHT schema that would help preserve the decentralized nature of Ethereum while preserving its resilience.

\section{Methodology}
\label{sec:methodology}
Ethereum's DAS approach presents an ideal yet ambitious synergy with DHTs. These networks offer light and resilient logical links across participants that can resist the sometimes chaotic dynamics of peer-to-peer networks. However, its direct application in such a demanding protocol as DankSharding remains unknown.
This section introduces the methodology we've used to simulate a DHT network with most parameters involved in a peer-to-peer network and under a heavy workload. Furthermore, we also introduce the developed framework that helped us validate the results of the simulations on the live IPFS network.   

\begin{figure*}
    \minipage{0.48\textwidth}%
        \centering
        \includegraphics[width=0.95\linewidth]{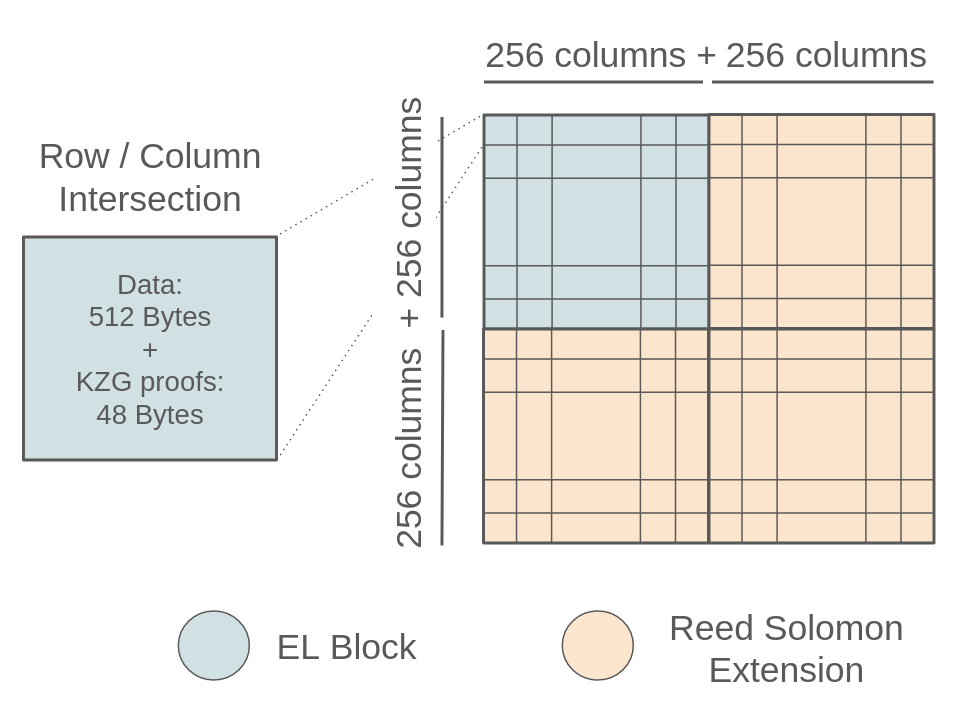}
        \caption{Block split proposal for DankSharding's DAS.}
        \label{fig:das-block-split}
    \endminipage\hfill
    \minipage{0.48\textwidth}%
        \centering
        \includegraphics[width=0.9\linewidth]{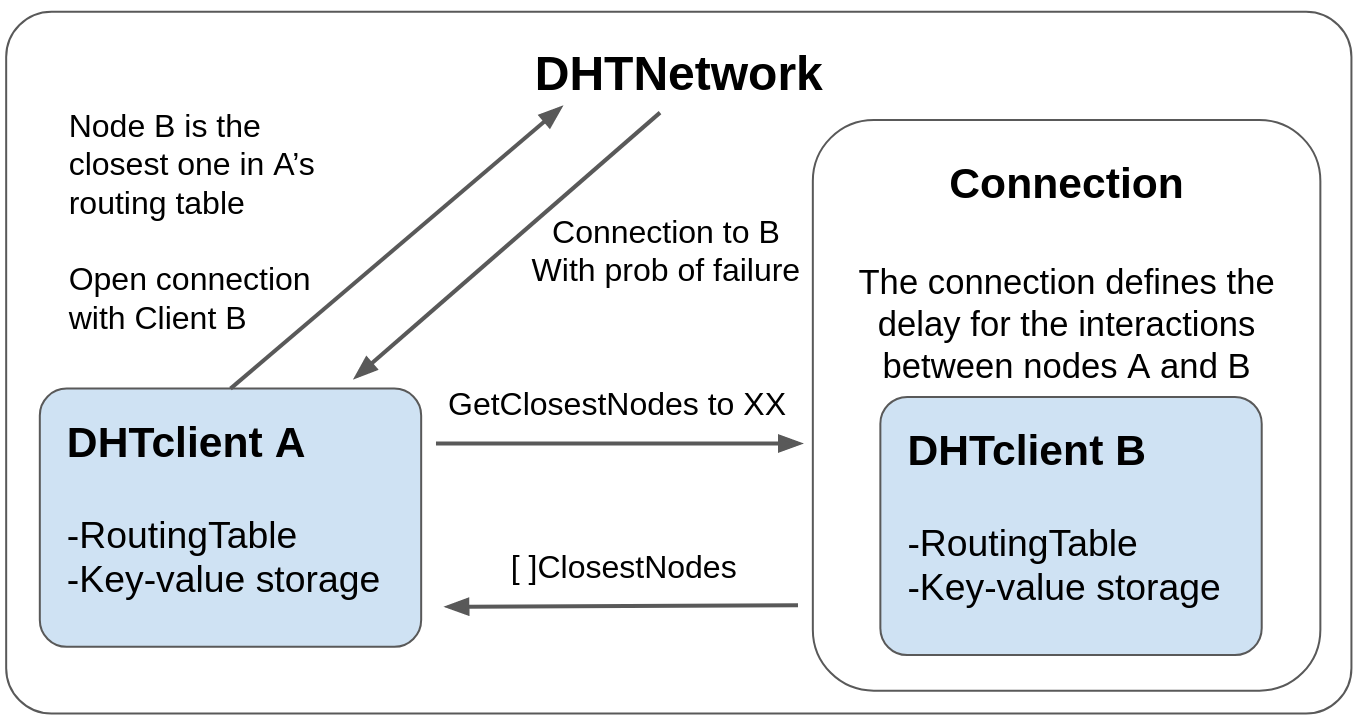}
        \caption{Description of the DHT components in the \emph{dht} simulator, and an example of the internal logic of their interaction.}
        \label{fig:py-dht}
    \endminipage
\end{figure*}

\subsection{DHT Simulator}
\label{subsec:simulator}
Since the internal parameters of the possible DHT are not defined (at the moment of writing this paper), the paper explores a possible DHT implementation for DAS through a simulator. The DAS simulator relies on a Python module that implements all the components of a Kademlia DHT called ``py-DHT"~\cite{py-dht}. The module allows the creation of a network of the given size, initializing the routing table of the nodes while using the XOR binary distance between nodeIDs to fill up the K-buckets. Inside the module, the two main components, the DHT network and the DHT nodes (also known as DHT clients), have different purposes. 

The network module offers a common shared perspective for the simulation, as shown in Figure \ref{fig:py-dht}. It keeps track of all nodes and simulates the communication between two nodes by introducing latencies and eventual losses. Because peer-to-peer networks tend to be subject to node congestion, node churn, or even connection churn \cite{henningsen2020crawling} \cite{daniel2022passively}, the network configuration allows to define the following set of parameters that helps reproduce any network behaviour:
\begin{itemize}
    \item Number of nodes participating in the DHT network.
    \item Fast error rate: ratio of connections that remote peers will refuse.
    \item Slow error rate: ratio of the connections that timeout when connecting a remote peer. 
    \item Connection delay range: latencies applied to the communication between two nodes.
    \item Fast delay range: latencies applied when a \emph{fast error} occurs.
    \item Slow delay range: latencies applied when a \emph{slow error} occurs.
    \item Gamma: the incremental delay applied to the communication between nodes when a peer is contacted by others, simulating the overhead of handling multiple connections.
\end{itemize}
On the other hand, the DHT clients are the ones that populate the simulated network. They include all the logic of the DHT, allowing themselves to swap information if the connection through the network interface succeeds. The simulator recreates the performance of a DHT, performing events such as initialization of the routing table, updating the routing table, looking up the closest peers to a key, providing a given value to the network, lookup for a specific key. The DHT clients have the configuration of the Kademlia DHT, including the following parameters:
\begin{itemize}
    \item K: replication factor for the DHT and k-buckets' size. 
    \item Alpha: maximum number of concurrent connections a DHT client can have while looking for a key/value or the closest clients to a key.
    \item Beta: number of close clients a remote peer provides when looking for a key.
\end{itemize}
We refer the reader to \cite{maymounkov2002kademlia} for the exact definition of these parameters.
The \emph{dht} [reference removed for double-blind review] module extracts all the timing and accuracy metrics from each individual operation, allowing the inspection of the network's performance under certain circumstances.

\subsection{Verification of the DHT simulations}
\label{subsec:verification-dht}
To validate the simulated results of the \emph{dht} module, the paper includes a comparison with the IPFS live network's performance. The results were validated by comparing the performance of the most basic operations in a DHT, such as DHT provides and DHT lookups in IPFS, with simulations replicating the same DHT and network parameters: $K=20$, $Alpha=3$, and $Beta=20$.
To obtain those performance measurements from the IPFS network, the team developed a dedicated tool (``looking-up-ipfs"~\cite{looking-up-ipfs}), which provides measurements of the internal operations of an IPFS's DHT-host interacting with the live network. 
The tool can concurrently generate and provide a given number of random Content Identifiers (CIDs) to the network\footnote{NOTE: when advertising content to the IPFS network, nodes only replicate the pointer to who has the content (Provider's Records) to the $k$ closest peers to the CID's hash values.}, dumping the duration of the process and the contacted peers while finding the $K$ closest peers of each content-provide operation.
Once the CIDs are published, the tool tests the retrievability of the same CIDs later, tracking the individual duration and the total contacted nodes during each independent operation. Furthermore, as it attempts the retrieval concurrently, it also allows us to measure the overhead of performing multiple low-demanding operations from a single DHT host.

\subsection{Development of the experiments}\label{subsec:hardware-for-studies}
Each experiment presented in the following evaluation Section~\ref{sec:evaluation} was conducted on dedicated hardware machines. The first part, the simulation of the DHT for Ethereum's DAS needs~\cite{das-research-dht}, was run on a PC with a Ryzen 5900X CPU, 32GB of memory, and 1TB of SSD disk. On the other hand, the IPFS benchmarks using the ``looking-up-ipfs" tools were run on an AWS cloud machine in Paris with 4 CPU cores, 8GB of memory, and 50GB of SSD disk on the 3rd of October, 2023.

\section{Evaluation}
\label{sec:evaluation}

\begin{figure*}
    \minipage{0.48\textwidth}%
        \centering
        \includegraphics[width=\linewidth]{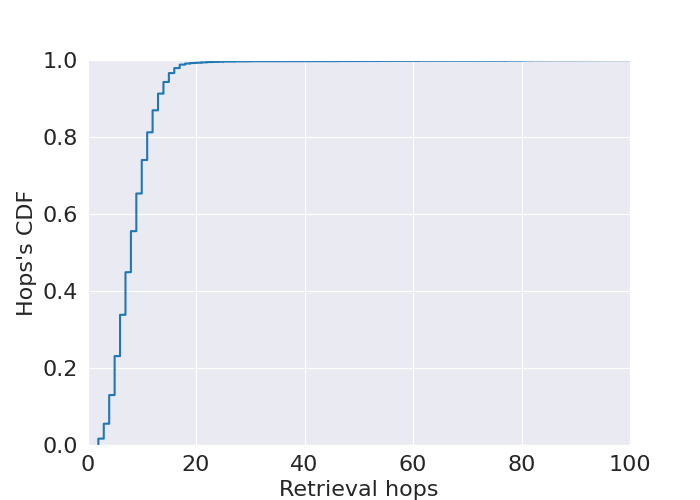}
        \caption{CDF of the DHT hops while performing 100 concurrent retrievals in the IPFS network.}
        \label{fig:ipfs-hops}
    \endminipage\hfill
    \minipage{0.48\textwidth}%
        \centering
        \includegraphics[width=\linewidth]{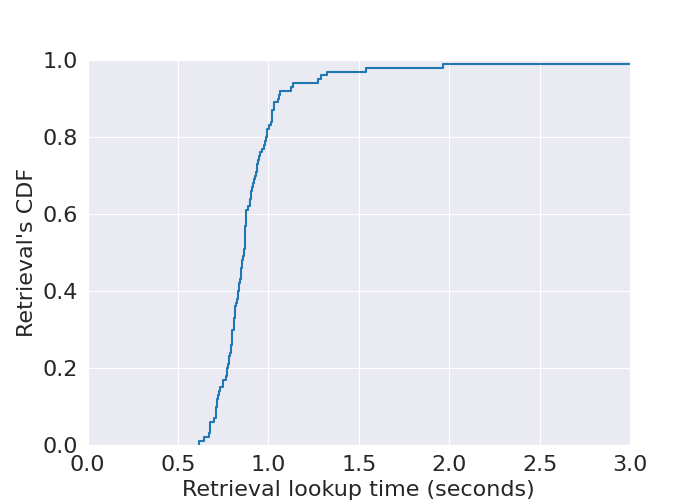}
        \caption{CDF of the time to finish $100$ sets of $80$ concurrent retrievals from the IPFS network.}
        \label{fig:ipfs-concurrent-retrievals}
    \endminipage
\end{figure*}

\subsection{DHT lookups}
\label{sebsec:lookups}

One of the key operations in a DHT is its ability to find which node in the network has the value for a specific key. This operation is generally known as \emph{DHT lookup}, and it consists of a set of recursive operations that, using the XOR distance between the node IDs and the given key, discover closer and closer nodes until, eventually, a node that has the value is found (assuming some node has it). DAS needs fast data retrievability from the DHT to ensure that blobs can be verified ``fast enough"\footnote{There is an ongoing debate in the Ethereum community about how fast this verification should be. It is commonly argued that one slot time (12~seconds) is fast enough.}. For this reason, measuring DHT's performance is critical to check if it would be a viable option for DAS in Ethereum. A peculiarity of DAS verification is that a node is starting concurrent lookup queries for several randomly selected samples. The DAS verification is considered complete when every single selected sample is retrieved. This section describes the main findings observed when simulating such concurrent queries on a DHT. We use $80$ queries from a node, as finding $80$ retrievable samples provides high enough probabilistic guarantees on a 512-by-512 segment 2D encoded block~\cite{das-on-danksharding}. 
We also run similar concurrent queries on the live IPFS network. This serves as a validation for the correctness of the simulator, where the number of hops and the aggregated delay for the operation match a live network if the parameters are correctly set. 

\begin{figure}
    \centering
    \includegraphics[width=0.65\linewidth]{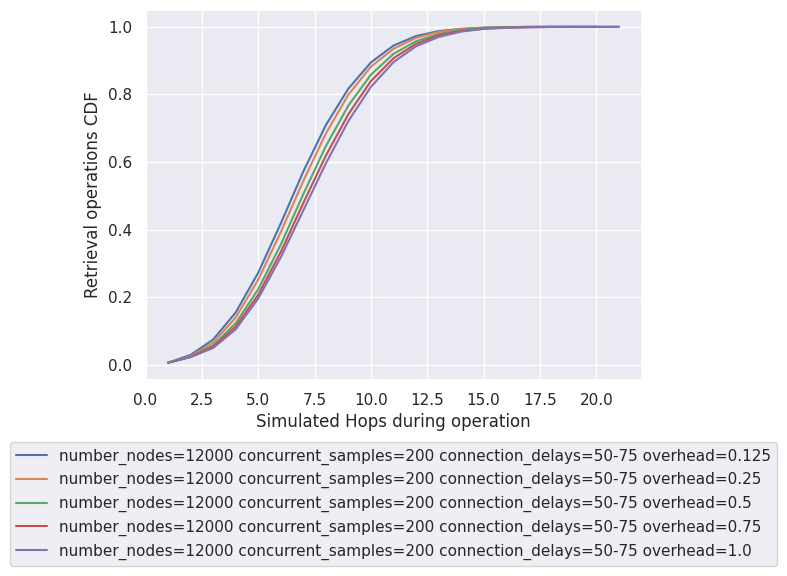}
    \caption{CDF of the simulated number of hops while looking for $200$ concurrent keys with different overheads with $k=20$, $alpha=3$, and $beta=20$.}
    \label{fig:simulated-hops}  
\end{figure}

\paragraph{Lookup hops}
\label{paragraph:lookup-hops}
A hop is an individual step in the recursive lookup operation. It implies asking a remote node in the network for a number ($beta$) of the closest peers it has to the given key or the value of the key if it has it. This is how the DHT narrows down the lookup of a value. Figure \ref{fig:ipfs-hops} shows the CDF of the number of hops done by an IPFS DHT node when performing $100$ sets of $80$ concurrent DHT lookups. The figure shows how the $99$\% of lookups were done in under $18$ hops, with a small tail in the last $1$\% that can reach the $100$ hops on some rare occasions. The numbers measured with the DHT simulator report similar results, as Figure \ref{fig:simulated-hops} shows. With a fast-failure rate of $10$\% of the connections, the simulation of $100$ sets of $200$ concurrent lookups report a $99$th percentile of $12$ to $14$ hops. 
We also notice that the concurrency overhead parameter does not impact the number of hops, as we see a negligible difference between the distributions for different overhead values. This makes sense as the time it takes a node to answer does not change the content of its routing table, and, therefore, the number of hops is not impacted. 


\begin{figure}
    \centering
    \includegraphics[width=0.85\linewidth]{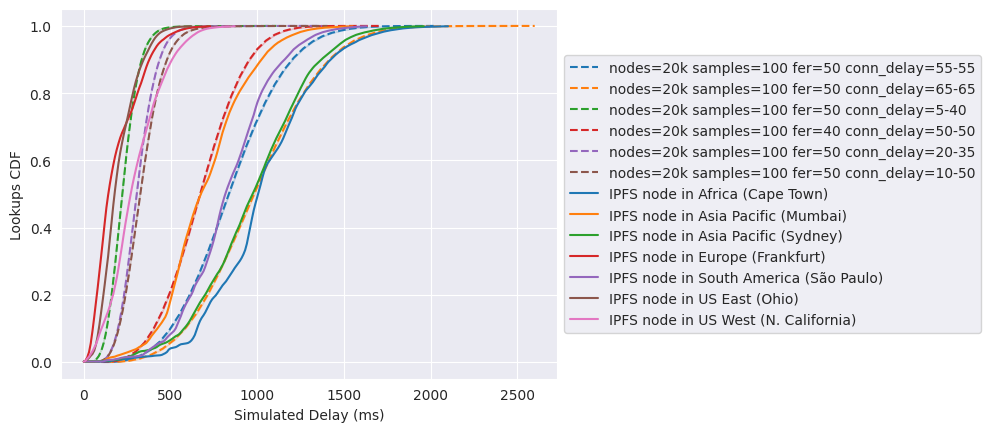}
    \caption{Accuracy of the DHT simulator model matching lookup performances in the IPFS network from different locations while keeping $k=20$, $alpha=3$, and $beta=20$. Note that ``fer" stands for ``fast error rate".}
    \label{fig:dht-validation-per-region}
\end{figure}

\paragraph{Time performance}
\label{paragraph:lookup-times}
Nodes join and leave the network as they please in permission-less networks such as IPFS and Ethereum. However, the users in the Ethereum network have shown more stability in maintaining their nodes for long periods. Users are generally staking or extracting on-chain data from nodes, so they are somehow incentivised to keep the nodes up and running. As a result, this stability can create healthier routing tables and potentially increase the DHT lookup time performance.
In an attempt to validate the impact of the network conditions on the DHT lookup performance, Figure \ref{fig:dht-validation-per-region} compiles the lookup time's CDF in the IPFS network done by the Probelab Team \cite{probelab_lookup_performance}, with the respective estimated network conditions in each of the regions. It is not a surprise that different regions have different connectivity. The difference on the $90$th percentile of the regions varies from $700$ milliseconds to retrieve a value in the US and Europe to $1.7$ seconds and $1.5$ milliseconds in South America and Africa. Moreover, this connectivity difference can magnify in p2p networks where nodes cluster more in certain regions. 
The figure also showcases the possibility of configuring the \emph{dht} module to fit any realistic network conditions. The figure shows that with the variation of the delay range at the simulated network, the results can accurately fit the measurements done in the IPFS DHT network from the lookup node's perspective. 

\begin{figure}
    \centering
    \includegraphics[width=0.65\linewidth]{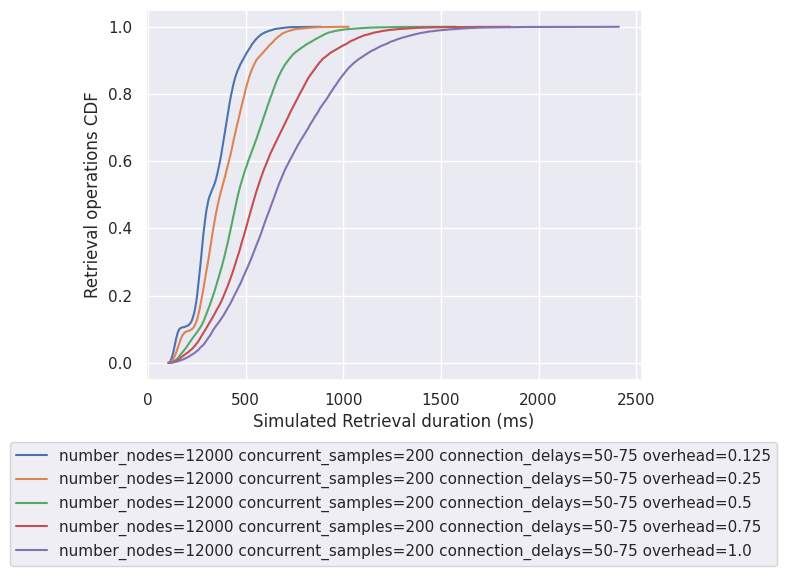}
    \caption{CDF of the simulated $100$ DHT retrieval sets with $200$ concurrent keys each using different overheads. The DHT simulation had the following parameters: $k=20$, $alpha=3$, and $beta=20$.}
    \label{fig:simulated-retrieval-time}  
\end{figure}

\paragraph{Concurrency overhead}
\label{paragraph:lookup-overhead}
We've mentioned the differences between the IPFS and the Ethereum network. The nature of IPFS, where $80$\% of users altruistically spawn a node and disconnect it after $8$ hours \cite{ipfs-churn}, makes it more subjected to node churn \cite{daniel2022passively}. This generally impacts the performance, as these nodes might be present in the routing table of other nodes for several minutes. Despite the expected node churn in Ethereum is expected to be lower, having non-active nodes inside the nodes' routing table doesn't only affect the direct time performance of the DHT operations. The main solution to overcome a certain connection failure rate is to attempt multiple simultaneous ones, with the extra cost of handling those extra concurrent requests. 
The presented $alpha$ parameter helps overcome the time impact of facing unreachable nodes, which in IPFS' is set to $3$ concurrent connections. However, it multiplies the cost of making a single lookup. Figure \ref{fig:simulated-retrieval-time} shows the time impact of different concurrency overhead parameters on $200$ concurrent DHT lookups from a single retrieving node's perspective. The overhead is applied to nodes contacted during the concurrent simulated operations. This means that each node connected during the lookup operation will apply an only increasing overhead delay to its connection delay as other concurrent operations contact the same peer. If a set $200$ connection needs to contact peer $A$ on three occasions, the first won't have any overhead delay, the second one will have a penalty of the overhead parameter, and the third one will have a penalty of two times the parameter. 
Users with access to larger hardware resources could be less restricted by the concurrency overhead of retrieving multiple block segments in sub-second lookups. In other occasions where the hardware resources are more limited, like in a solo-staker scenario, the DAS operations will be more penalized for handling so many interactions. This exposes a clear disadvantage for smaller users if the number of lookups is too high in the long-term projection of the protocol.

To compare the existing overhead in the IFPS DHT client, Figure \ref{fig:ipfs-concurrent-retrievals} exposes the CDF time of concluding $80$ concurrent lookups in the IPFS network in a set of $100$ rounds. The figure shows that $90$\% of concurrent lookups performed between $600$ms and $1.2$ seconds. 
Compared to the numbers presented in Figure \ref{fig:dht-validation-per-region}, we can clearly see that IPFS's client does see an impact in performance originating from a concurrency overhead, not at least at this number of concurrent requests. Therefore, it is also an expected effect in a possible DHT for Ethereum's DAS.

\subsection{Seeding of the DHT}
\label{subsec:provides}
The results of the experiments are clear: modern Kademlia-based DHTs can achieve sub-second sampling performance on its $90$th percentile with the favourable conditions that the Ethereum network provides. Even if looking at less favourable conditions and at higher percentiles, we see latencies well below the slot time of $12$ seconds.
However, despite the logarithmic scaling solution for content addressing, the idea of the DHT seems to be reaching its limits in highly demanding conditions. 
While the sampling itself seems to work, the first step of a DHT solution is seeding the block data into the DHT, where the samples are dispersed around the network according to their position in the DHT address space.
\begin{figure*}
    \minipage{0.49\textwidth}%
        \centering
        \includegraphics[width=\linewidth]{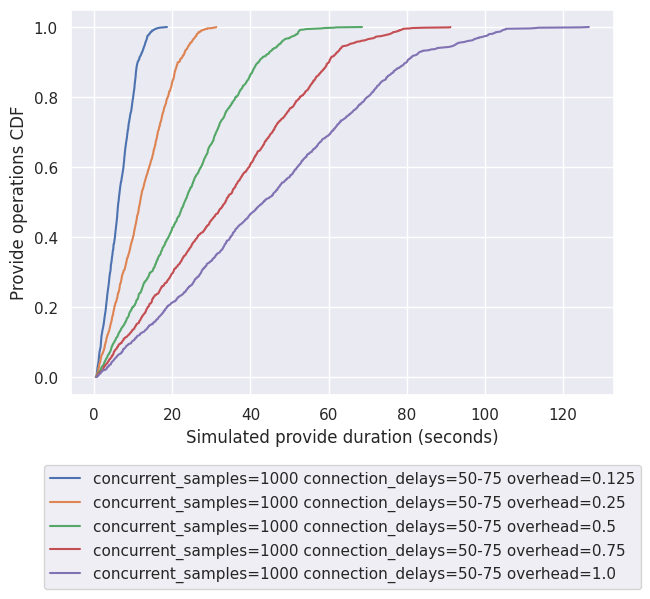}
        \caption{CDF of the simulated $1.000$ concurrent DHT provide operations from a single node.}
        \label{fig:simulated-1k-provides}
    \endminipage\hfill
    \minipage{0.49\textwidth}%
        \centering
        \includegraphics[width=\linewidth]{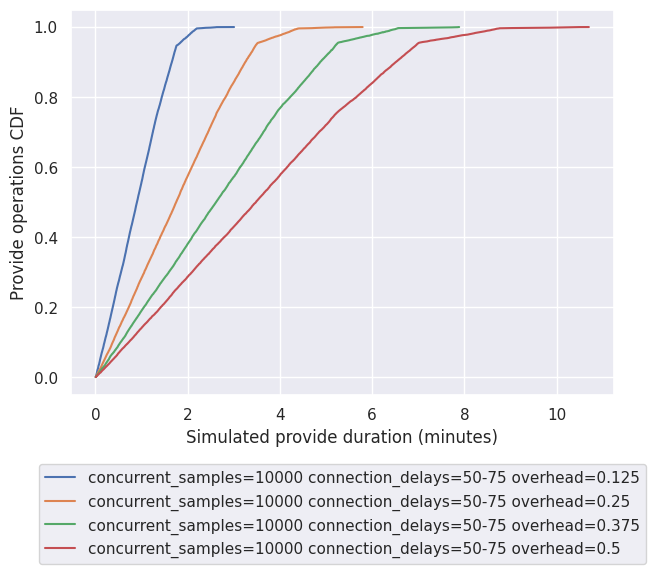}
        \caption{CDF of the simulated $10.000$ concurrent DHT provide operations from a single node.}
        \label{fig:simulated-10k-provides}
    \endminipage
\end{figure*}
Ethereum's current network conditions report roughly $13.000$ active nodes. Combined with the suggested standard Kademlia parameters \cite{pr-liveness} of $k=20$ and $alpha=3$, the seeding of $262.144$ block segments (512 columns * 512 rows) would produce a ratio of $403.29$ block segments that each node in the network would have to store, on average, every $12$ seconds. 
Moreover, storing the block samples doesn't only affect the last contacted nodes. Providing the keys is a process that starts from a distinct lookup to find the closest $k$ nodes to the segment's key, performing a median of $8$ to $10$ hops until the closest nodes are notified.
If handled simply as a large number of individual storage requests, the overall network load can easily overload the network, considering that all this should happen quickly before the actual sampling begins.

To simulate how tedious this seeding from a single node step would look in a Kademlia DHT, Figure ~\ref{fig:simulated-1k-provides} and ~\ref{fig:simulated-10k-provides} show the CDF of the simulated DHT provide times for $1.000$ and $10.000$ samples concurrently in a network of $12.000$ nodes. They show that the overhead factor, together with the value of $k=20$, has a huge impact on the total duration of the operation. Figure~\ref{fig:simulated-1k-provides} shows that a low number of samples, $1.000$, can still be broadcast to the network within $20$ seconds. However, it can only be achieved if the overhead generated in the network remains low, and existing users would need to upgrade their hardware to ensure enough resources to keep the DHT. As we increase the number of concurrently distributed samples from a single node (check ~\ref{fig:simulated-10k-provides} with $10.000$ concurrent samples), the CDF shows that the DHT starts to find a bottleneck. 

\begin{figure*}
    \minipage{0.49\textwidth}%
        \centering
        \includegraphics[width=\linewidth]{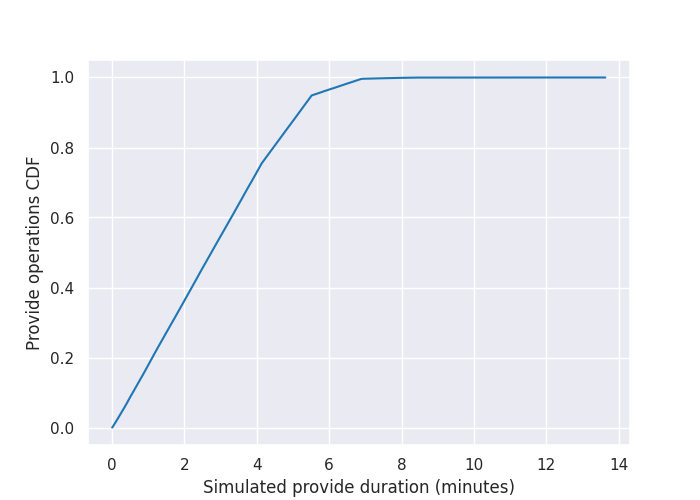}
        \caption{CDF of the simulated $262.144$ concurrent DHT provide operations from a single node.}
        \label{fig:simulated-262k-provides}
    \endminipage\hfill
    \minipage{0.49\textwidth}%
        \centering
        \includegraphics[width=\linewidth]{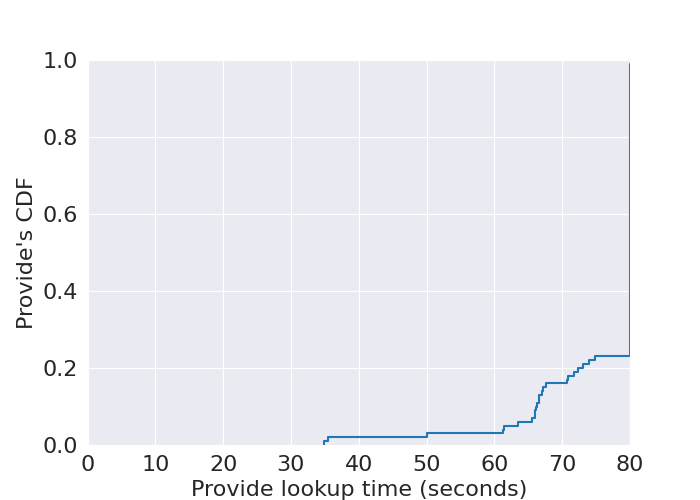}
        \caption{CDF of the time to finish $100$ sets of $80$ concurrent CID provides to the IPFS network.}
        \label{fig:ipfs-concurrent-provides}
    \endminipage
\end{figure*}

The biggest bottleneck comes from the routing nodes on the seeder's routing table. As Kademlia DHT keeps the most optimal and active nodes in the routing table using the XOR distances, the routing table barely changes in such a short time. Thus, those $250$ to $300$ in the routing table are always contacted first on each provided operation, being the most overwhelmed nodes in the network. This gets more visible if we simulate the provided operation from a single node (i.e., the block builder\cite{proposer-builder}) of the $262.144$ samples inside a blob block. Figure~\ref{fig:simulated-262k-provides} shows that even in a simulation of the best possible conditions (setting a very low factor of $0.015$ms per connection), the congestion in the network would make the process delayed to the $10$ to $14$ minutes, existing $50$ times the original broadcasting deadline of the $12$ seconds.         
To correlate these simulated findings with a real Kademlia DHT implementation, Figure~\ref{fig:ipfs-concurrent-provides} shows the CDF of the total duration of providing $100$ sets of $80$ CIDs in the IPFS network (from the beginning of the first CID provide operation, to the end of the last one in a single set of $80$ CIDs). The figure shows that barely $20$\% of the hundred sets of provides concluded within $70$ seconds, while almost $67$\% of the provides sets were limited by the timeout of $80$ seconds for each provide.
When considering the network load of seeding the DHT, it is important to remember that block propagation is also done to validators using GossipSub. However, in that case, instead of sending individual samples to selected target nodes, the efficient mesh-based GossipSub protocol is used to send entire rows and columns. As mentioned before, serving sampling requests directly from validators would compromise the identity of individual validator nodes, and a default Kademlia DHT can't handle that level of request in such a short time range. Thus, this raises the question: how do we seed the DHT, and can a DHT still be used as a blob addressing protocol?


\subsection{DHT limitations}
\label{subsec:dht-limitations} 

The DHT could be seeded by the block proposer/builder, but also by validators (limited to the scope of their GossipSub subscription). However, we can state that both cases are problematic for several reasons explained in the following paragraphs:

\subsubsection{Single seeder to the DHT}
\label{par:single-seeding} 
If the block proposer/builder is in charge of the DHT seeding, the nodes in its routing table will suffer the biggest workload of replying to the first hops of the $262.144$ requests, creating a huge workload that the network can't handle in $12$ seconds as Figure~\ref{fig:simulated-262k-provides} shows. The tedious problem of having to perform many provide operations to the DHT network is already present in the IPFS DHT for large service providers like Cloudflare \cite{cloudflare} or Pinata \cite{pinata}.
As a result, organizations and developers have moved the biggest part of the content in IPFS away from the DHT, relying on several solutions:
\begin{itemize}
    \item BitSwap~\cite{de2021accelerating}, a protocol designed to download content from a connected network participant optimally. Using BitSwap connections as a solution avoids seeding the ``provider records" for each hosted item. However, it relies on the nature of the network to end up connected to the largest DHT nodes, which centralizes the network and doesn't provide any content-addressing solution. It relies on the probability of eventually being connected to one of these ``big" nodes to retrieve the desired content, not guaranteeing minimal retrieval times for a non-popular item. 
    \item Network Indexers\footnote{https://github.com/ipni}. 
    It is a separate method that tackles the scalability problems of the DHT and the content addressing limitations of BitSwap. It offers a network of high throughput and low latency DHT-like databases present in the network as nodes that can be used by users supporting the protocols. As expected, this solution breaks the nature of distributed networks, as it relies on adding a particular actor whose only function is to provide the content addressing in the network. The solution, despite being scalable, magnifies the exposition of the network to a denial of service attack. Thus, it is not a viable option for Ethereum.
    \item The Accelerated Routing Table\footnote{https://github.com/ipfs/kubo/pull/8997/files}.
    This experimental routing table enables the possibility of making batch provides or reprovide sweeps\footnote{https://github.com/libp2p/go-libp2p-kad-dht/issues/824}.
    The method relies on the crawling capabilities of the DHT host to discover all the participants in the network. Based on the whole pack of content it has to advertise to the network, it can dismiss every DHT lookup to discover the $K$ closest peers as it already knows the entire list of participants. As a result, a node using an accelerated routing table can perform a bulk-provide operation per each node it needs to connect. The method saves many round-trip connections but exposes the network to the feared node churn. One could think about increasing the frequency of refreshing the routing table to overcome the node churn; however, it would also mean crawling the network more often, replacing the network's spam from DHT lookups and refreshing the accelerated routing table. Furthermore, the Ethereum network tends to have a low level of connections to limit the bandwidth usage~\cite{cortes2021resource}, which makes it unviable to implement at the moment.
    \item Recursive (and forwarding) Kademlia. The cost of seeding a large number of small segments (key-value pairs) into a Kademlia DHT is expensive due to the large number of interactions needed from the iterative lookup process, the parallel lookups (alpha), and the number of target storage nodes (K). \cite{heep2010rkademlia} introduced recursive Kademlia to reduce the iterative overhead, while \cite{czirkos2013, kadcast} extended this to support efficient broadcasting in the Kademlia address space. While DAS seeding is more complex than broadcasting, and recursive forwarding introduces compromises in the robustness of seeding the DHT; similar solutions can be considered to reduce the network load and time required. 
\end{itemize} 

\subsubsection{Multiple seeders to the DHT}
\label{par:multiple-seeding} 
On the other hand, the seeding could be done more organically by peers subscribed to a small set of GossipSub topics. This would balance the workload of publishing many items to the DHT for individual routing tables of the seeding nodes. But even with more sophisticated solutions like Optimistic Provide~\cite{trautweinipfs}, the network ends up over-flooded by the same message as more nodes perform the same provide operation. This solution could be improved by adding some logic to the consensus, where validators rely on deterministic randomness to define who must seed the DHT. This solution could follow a similar approach to the one that defines which validator aggregates and distributes the attestations at each slot committee. However, it requires a heavy change in the current logic of the consensus and could expose the privacy of the validators to some degree.

\subsubsection{Avoiding the DHT}
Finally, we should mention that while writing this paper, some early-stage emerging proposals try to avoid using a DHT for sampling. PeerDAS~\cite{peer-das} relies on a deterministic segment-to-nodeID mapping scheme to avoid expensive lookups. Sampling nodes are assumed to know enough node IDs to be able to perform their sampling based on the mapping. While a peer sampling service is assumed, the compromise of PeerDAS compared to the DHT-based solution is a reduced number of rows and columns.
Another proposal, SubnetDAS~\cite{subnet-das}, instead compromises on the randomness of the sampling, more specifically on the unlinkability of the queries. While in a DHT-based design every single node can query its selected random samples, in SubnetDAS nodes subscribe to GossipSub channels, and hold these subscriptions for a specified period, weakening the protection of nodes against double-spend.

\section{Conclusion}
\label{sec:conclusion}

This paper presents the results and limitations of implementing a direct Kademlia DHT as a DAS solution for Ethereum's scalability proposal. We offer a DHT simulator that is validated using experiments in a production p2p network with an actual Kademlia DHT implementation. Our results expose that, despite being the right fit for data sampling, it is problematic to seed the DHT in the context of Ethereum DAS. We discuss several possible alternative ways to seed the DHT, as elements of other designs not relying on a DHT. 
In future work, we intend to explore the viability and performance of DHT solutions in the context of more flexible and less constrained DAS protocols.


\printbibliography

\end{document}